\documentclass[aps, prc, reprint, twocolumn, amsmath, amssymb, superscriptaddress, floatfix, nofootinbib, notitlepage]{revtex4-1} 
\usepackage[pdftex]{graphicx}
\DeclareGraphicsExtensions{.jpg,.png,.pdf}
\usepackage[utf8]{inputenc}
\usepackage{slashed}
\usepackage{hyperref}
\usepackage{amsmath}
\usepackage{amssymb}
\usepackage{amsfonts}
\usepackage{tabularx}
\usepackage{booktabs}
\usepackage{graphicx}
\usepackage{color}
\usepackage{multirow}
\usepackage[perpage,symbol]{footmisc}
\usepackage{verbatim}
\usepackage{dcolumn}
\usepackage{bm}
\usepackage[inline]{enumitem}
\graphicspath{{Fig/}}
\definecolor{theblue}{RGB}{0,50,230}
\hypersetup{
	colorlinks=true,
	linkcolor=theblue,
	citecolor=theblue,
	urlcolor=theblue
}

\usepackage{lineno}

\begin{document}

\title{
Heavy-quark transport across the QCD crossover driven by a lattice-constrained in-medium potential
}

\author{Wu~Wang}
\thanks{These authors contributed equally to this work.}
\affiliation{College of Mathematics and Physics, College of Nuclear Energy Science and Engineering, China Three Gorges University, Yichang 443002, China}

\author{Yuqi~Luo}
\thanks{These authors contributed equally to this work.}
\affiliation{College of Mathematics and Physics, College of Nuclear Energy Science and Engineering, China Three Gorges University, Yichang 443002, China}

\author{Fei~Sun}
\affiliation{College of Mathematics and Physics, College of Nuclear Energy Science and Engineering, China Three Gorges University, Yichang 443002, China}
\affiliation{Center for Astronomy and Space Sciences, China Three Gorges University, Yichang 443002, China}

\author{Sa~Wang}
\affiliation{College of Mathematics and Physics, College of Nuclear Energy Science and Engineering, China Three Gorges University, Yichang 443002, China}
\affiliation{Center for Astronomy and Space Sciences, China Three Gorges University, Yichang 443002, China}

\author{Jungang~Deng}
\affiliation{College of Mathematics and Physics, College of Nuclear Energy Science and Engineering, China Three Gorges University, Yichang 443002, China}
\affiliation{Center for Astronomy and Space Sciences, China Three Gorges University, Yichang 443002, China}

\author{Wei~Xie}
\affiliation{College of Mathematics and Physics, College of Nuclear Energy Science and Engineering, China Three Gorges University, Yichang 443002, China}
\affiliation{Center for Astronomy and Space Sciences, China Three Gorges University, Yichang 443002, China}

\author{Shuang~Li}
\email[Contact author: ]{lish@ctgu.edu.cn}
\affiliation{College of Mathematics and Physics, College of Nuclear Energy Science and Engineering, China Three Gorges University, Yichang 443002, China}
\affiliation{Center for Astronomy and Space Sciences, China Three Gorges University, Yichang 443002, China}

\author{Kejun~Wu}
\email[Contact author: ]{wukj@ctgu.edu.cn}
\affiliation{College of Mathematics and Physics, College of Nuclear Energy Science and Engineering, China Three Gorges University, Yichang 443002, China}
\affiliation{Center for Astronomy and Space Sciences, China Three Gorges University, Yichang 443002, China}

\date{\today}

\begin{abstract}
We present a self-consistent framework for heavy-quark transport in the quark-gluon plasma across the QCD crossover region. By synthesizing perturbative and nonperturbative interactions into a unified interaction kernel, we circumvent the traditional reliance on arbitrary soft-hard momentum separation scales. The interaction is governed by an in-medium effective potential, incorporating short-range Yukawa screening and long-range confining string contributions, both rigorously constrained by the latest lattice QCD data. Our results reveal that the nonperturbative string tension is indispensable for capturing the extreme opacity of the medium near the critical temperature $T_c$. Specifically, our model predicts a spatial diffusion coefficient of $2\pi T D_s \approx 0.5 \sim 1.7$, demonstrating a striking quantitative agreement with the recent lattice QCD extractions. Ultimately, our results provide a robust dynamical interpretation of the strong heavy-quark coupling near the QCD crossover and offer a unified framework for describing heavy-flavor transport in hot and dense QCD matter.
\end{abstract}

\maketitle

\section{Introduction}\label{sec:introd}

Heavy quarks (charm and bottom) serve as exceptionally well-controlled probes of the strongly interacting quark-gluon plasma (QGP) produced in ultrarelativistic heavy-ion collisions. Owing to their large masses ($m_{Q} \gg T_{\rm QCD}$), they are predominantly produced in initial hard scatterings and subsequently propagate through the dynamically evolving medium. Consequently, they encode essential information about the medium's microscopic transport properties and color screening structure~\cite{He:2022ywp, Apolinario:2022vzg, Chen:2021akx, Zhao:2020jqu, Cao:2020wlm, RalfSummary16}.

In previous works~\cite{Lou:2025wmw, Peng:2024zvf, Li:2021nim}, heavy-quark interactions in a thermal medium were typically described within a soft-hard factorization framework. In this approach, soft momentum transfers are treated using hard thermal loop (HTL) resummed propagators, while hard scatterings are evaluated utilizing bare perturbative matrix elements. These two regimes are conventionally separated by an intermediate momentum scale $\sqrt{-t^\ast}$. Although this scheme provides a rigorously controlled weak-coupling baseline, its reliance on an explicit separation scale introduces unphysical dependencies, which become increasingly problematic in the strongly coupled regime near the QCD crossover temperature $T_c$.

Lattice QCD calculations consistently indicate that near $T_c$, the medium develops strong nonperturbative chromoelectric correlations that drastically deviate from weak-coupling expectations. As a consequence, purely perturbative approaches tend to severely underestimate the interaction strength, leading to spatial diffusion coefficients ($2\pi T D_s$) that are substantially larger than lattice-extracted values~\cite{Pisarski:2000eq, He:2022ywp}. To remedy this, various phenomenological approaches have been proposed, including chromo-magnetic-monopole~\cite{Liao:2006ry, Liao:2008jg, Liao:2008dk}, quasiparticle models (QPM)~\cite{PhysRevD.54.2399, Peshier:2002ww, Plumari:2011mk, Sambataro:2025obe, Grishmanovskii:2025mnc, Krishna:2025bll, Sambataro:2024mkr}, background-field-based descriptions~\cite{Peng:2025xwe, Du:2024riq}, and nonperturbative $T$-matrix frameworks~\cite{vanHees:2007me, He:2011qa}. The latter particularly emphasizes the resummation of the interaction potential to account for the emergence of resonant correlations and potential bound states in the crossover region. However, a persistent limitation of existing frameworks is the absence of a direct, systematic mapping between lattice-constrained static potentials and the real-time scattering amplitudes governing heavy-quark transport without invoking arbitrary cutoffs.

In this work, we develop a unified approach to bridge this gap. Our theoretical starting point is the in-medium heavy-quark potential constructed within the generalized Gauss-law framework, which explicitly incorporates both Coulombic and confining string interactions. By calibrating this potential against lattice QCD data, we extract an effective screening mass $M_{\rm D}(T)$. We then perform an analytical Fourier transform to map the coordinate-space potential onto a momentum-space interaction kernel. This kernel is directly implemented as the effective $t$-channel propagator in heavy-quark scattering processes, enabling a continuous description across the full momentum range. By assigning appropriate Lorentz structures to the interaction vertices, this framework ensures perturbative QCD consistency at high momentum transfers while naturally incorporating lattice-driven nonperturbative enhancements in the infrared region.

The remainder of this paper is organized as follows. In Sec.~\ref{sec:potential}, we introduce the lattice-constrained generalized Gauss-law potential at zero density. In Sec.~\ref{sec:amplitudes}, we construct the corresponding unified dynamical scattering amplitudes. The formulation of heavy-quark transport coefficients is presented in Sec.~\ref{sec:transport}. Numerical results are discussed in Sec.~\ref{sec:results}, followed by a summary in Sec.~\ref{sec:summary}.

\section{Lattice-Constrained In-Medium Effective Potential}\label{sec:potential}

To overcome the limitations of the perturbative baseline near $T_c$, we introduce a nonperturbative extension that modifies the infrared structure of the interaction kernel while preserving perturbative consistency in the ultraviolet sector. 
The construction is based on the generalized Gauss law potential constrained by lattice QCD.

\subsection{Effective screening mass}

To quantify nonperturbative medium effects on heavy-quark interactions, the vacuum Coulomb potential alone~\cite{Eichten:1978tg,Eichten:1979ms} is insufficient. Instead, we employ the in-medium heavy-quark potential derived from the generalized Gauss law supplemented by the HTL permittivity~\cite{Rothkopf:2019ipj,GenGaus1990}. 
In this framework, the vacuum Cornell potential,
\begin{equation}\label{eq:V_vacu}
V_{\rm vac}(r,0)=-\frac{\tilde{\alpha}_s}{r}+\sigma r + V_0,
\end{equation}
consisting of a Coulombic part and a linear confining part, is embedded into a thermal medium through linear response theory. Here, $\tilde{\alpha}_s$ denotes an effective short-distance coupling parameter, which is treated phenomenologically alongside the string tension $\sigma$ and the constant shift $V_0$. 

In lattice QCD simulations, finite-temperature ensembles are characterized by specific lattice couplings $\beta$. While these vacuum parameters are often interpolated across lattice spacings to match thermal simulations~\cite{Lafferty:2019jpr}, implementing them into a unified dynamical scattering kernel requires a single, consistent set of temperature-independent baseline parameters. Therefore, we compute the weighted average of these parameters across the relevant lattice ensembles, utilizing their respective systematic uncertainties as weights. This procedure robustly fixes the baseline zero-temperature potential, yielding the central values:
\begin{equation}\label{eq:Params_vacu}
\tilde{\alpha}_s = 0.406, \quad \sqrt{\sigma} = 0.495~{\rm GeV}, \quad V_{0} = 2.356~{\rm GeV}.
\end{equation}

The real part of the in-medium potential reads as
\begin{equation}\label{eq:ReVtotal_new}
V(r,T)= V_{\rm Y}(r,T) + V_{\rm S}(r,T) + V_0,
\end{equation}
where the screened Coulombic (Yukawa) and string components are~\cite{Lafferty:2019jpr}, respectively,
\begin{equation}
\label{eq:ReVc_new}
V_{\rm Y}(r,T) = -\tilde{\alpha}_s \left( M_{\rm D} + \frac{e^{-M_{\rm D}r}}{r} \right),
\end{equation}
\begin{equation}
\label{eq:ReVs_new}
V_{\rm S}(r,T) = \frac{2\sigma}{M_{\rm D}} - \frac{\sigma}{M_{\rm D}} e^{-M_{\rm D}r} \left(2+M_{\rm D}r\right).
\end{equation}
The screening mass $M_{\rm D}(T)$ characterizes the spatial attenuation of the chromoelectric fields. As demonstrated in Eqs.~(\ref{eq:ReVc_new}) and (\ref{eq:ReVs_new}), this construction correctly preserves the vacuum limits: for $r\to0$, medium effects vanish, yielding $V_{\rm Y}\to -\tilde{\alpha}_s/r$ and $V_{\rm S}\to \sigma r$. Conversely, at large distances, the exponential factors effectively suppress both contributions, reflecting dynamic chromoelectric screening.

With the baseline vacuum parameters fixed, the effective screening mass $M_{\rm D}(T)$ is extracted by fitting the lattice data for $V(r,T)$ at finite temperatures~\cite{Burnier:2014ssa, Bazavov:2023dci}. The continuum-corrected screening mass is then parametrized using a robust HTL-inspired functional form~\cite{Lafferty:2019jpr, Burnier:2015nsa, Kajantie:1997pd}:
\begin{equation}\label{eq:mDEff_vsT_new}
M_{\rm D}(T) = m_{\rm D}(T) + \frac{N_c g^2 T}{4\pi} \ln\!\left[ \frac{m_{\rm D}(T)}{g^2 T} \right] + \kappa_1 g^2 T + \kappa_2 g^3 T,
\end{equation}
where $m_{\rm D}(T)$ denotes the leading-order HTL Debye mass at vanishing chemical potential,
\begin{equation}
\begin{aligned}\label{eq:MD_WithMu_HTL_mu0}
	m_{\rm D}^2(T) &= \biggr(\frac{N_c}{3}+\frac{N_f}{6}\biggr)g^2 T^2.
\end{aligned}
\end{equation}
Here, $N_c = 3$ is the number of colors for $SU(N_{c})$ symmetry
and $N_f = 3$ is the number of quark flavors.
The logarithmic term in Eq.~(\ref{eq:mDEff_vsT_new}) corresponds to the known next-to-leading-order correction~\cite{Arnold:1995bh, Rebhan:1993az}. 
The running coupling $g(\Lambda)=g(2\pi T)$ is evaluated using the four-loop QCD beta function in the $\overline{\rm MS}$ scheme~\cite{vanRitbergen:1997va,Czakon:2004bu,Chetyrkin:1997sg}, initialized with $\Lambda_{\rm QCD}^{(5)}=0.2145\,{\rm GeV}$ and matched across quark thresholds~\cite{Chetyrkin:1997un}.

Rather than performing a new baseline fit, we strictly adopt the robust central parametrization established in the literature. Specifically, we follow the rigorous methodology detailed in Refs.~\cite{Lafferty:2019jpr, Burnier:2015tda}. In these references, the extraction of $\kappa_1$ and $\kappa_2$ is strictly executed via a three-step procedure: First, the discrete lattice screening mass, $M_D^{\text{lat}}(T)$, is extracted by fitting the real part of the theoretical in-medium potential to the available finite-temperature lattice QCD static potential data. During this thermal fitting process, the lattice-scale vacuum parameters are strictly frozen to isolate pure medium-induced screening modifications and eliminate vacuum-related phenomenological ambiguities. Second, to systematically remove the unphysical ultraviolet cutoff artifacts (discretization errors) inherent in finite lattice spacings, these discrete masses are projected into the continuum limit. This is achieved via the dimensionless scaling relation $M_D^{\text{cont}}(t_{\text{red}}) = M_D^{\text{lat}}(t_{\text{red}}) \times \sqrt{\sigma^{\text{cont}}/\sigma^{\text{lat}}(\beta)}$ evaluated at identical reduced temperatures ($t_{\text{red}} = T/T_c$), where $\sigma^{\text{cont}}$ is the vacuum string tension obtained from the prior zero-temperature calibration, and $\sigma^{\text{lat}}(\beta)$ is the corresponding lattice-scale string tension derived via internal interpolation at a specific gauge coupling $\beta$~\cite{Burnier:2015tda}. Finally, the parametrized functional form of the effective Debye mass defined in Eq.~(\ref{eq:mDEff_vsT_new}) is fitted directly to these continuum-extrapolated values, $M_D^{\text{cont}}(T)$, within the temperature range from approximately $0.9T_c$ up to approximately $1.7T_c$. This uniquely determines the central coefficients:
\begin{equation}
\label{eq:KappaParams_new}
\kappa_1 = 0.686, \qquad \kappa_2 = -0.317.
\end{equation}

To rigorously quantify the theoretical uncertainties propagating from the nonperturbative magnetic sector into the heavy-quark transport coefficients, we independently evaluate the parameter boundaries. Instead of relying solely on the central values, we define two additional parameter sets representing the upper and lower boundaries of the effective Debye mass uncertainty band. The upper and lower boundary values of our coefficients, denoted as $(\kappa_1^{\text{upp}}, \kappa_2^{\text{upp}})$ and $(\kappa_1^{\text{low}}, \kappa_2^{\text{low}})$, are uniquely determined by matching our parametrization to the upper and lower limits of the continuum-extrapolated $M_D$ uncertainty bands reported in Ref.~\cite{Lafferty:2019jpr}. Specifically, we implement the values $\kappa_1^{\text{upp}} = 0.862$ and $\kappa_2^{\text{upp}} = -0.348$ to establish the upper bound constraint. Conversely, the lower bound constraint is established utilizing $\kappa_1^{\text{low}} = 0.421$ and $\kappa_2^{\text{low}} = -0.245$. This comprehensive procedure ensures that our interaction kernel and its associated uncertainty bands are strictly constrained by the state-of-the-art lattice QCD evaluations. Consequently, the transport coefficients calculated using these specific parameter sets formulate a reliable theoretical uncertainty band in our subsequent numerical results.

In the numerical implementation of the thermal scaling, we adopt a critical temperature of $T_c = 172.5$ MeV. This specific value is chosen to maintain strict consistency with the original lattice QCD studies of the static potential~\cite{Lafferty:2019jpr}, ensuring that the dimensionless temperature $T/T_c$ correctly maps the nonperturbative features observed in the lattice data.

To transparently summarize the baseline physical scales employed in our numerical calculations, we explicitly implement $m_c = 1.5$~GeV for charm quarks and $m_b = 4.75$~GeV for bottom quarks. We emphasize that we strictly adopt the pole mass scheme, which is highly consistent with the standard central values utilized in the fixed order plus next-to-leading logarithm (FONLL) calculations for heavy-flavor hadroproduction~\cite{Cacciari:2012ny}. Within our theoretical framework, the heavy-quark scattering matrix elements are formulated under a nonrelativistic approximation in the medium rest frame ($m_Q \gg T$). The pole mass scheme is the most consistent choice for this kinematic regime, as it appropriately defines the on shell kinematic boundaries and strictly justifies the static limit dispersion relation ($v \sim \sqrt{T/m_Q} \ll 1$).

\subsection{Momentum-space potential $\tilde{V}(q, T)$}
To evaluate the dynamical scattering cross sections, we map the coordinate-space interaction into momentum space via the Fourier transform: $\tilde{V}(\vec{q}, T) = \int d^3\vec{r} e^{-i\vec{q}\cdot\vec{r}} V(r, T)$. Executing this transform yields the full momentum-space effective potential:
\begin{equation}
\label{eq:Vq_Total}
\tilde{V}(q, T) = \tilde{V}_{\rm Y}(q, T) + \tilde{V}_{\rm S}(q, T) + (2\pi)^3 V_{\rm const} \delta^{(3)}(\vec{q}\,),
\end{equation}
where $q = |\vec{q}\,|$, and the individual components are analytically given by:
\begin{equation}\label{eq:Vq_Components_Y}
\tilde{V}_{\rm Y}(q, T) = -\frac{4\pi \tilde{\alpha}_s}{q^2 + M_{\rm D}^2},
\end{equation}
\begin{equation}\label{eq:Vq_Components_S}
\tilde{V}_{\rm S}(q, T) = -8\pi\sigma \frac{q^2 + 5M_{\rm D}^2}{(q^2 + M_{\rm D}^2)^3}.
\end{equation}

The coordinate-space potential includes a constant term $V_{\rm const} = V_0 - \tilde{\alpha}_s M_{\rm D} + 2\sigma/M_{\rm D}$ that encapsulates uniform background field contributions. The Fourier transform of this constant potential yields a Dirac delta function $\delta^{(3)}(\vec{q}\,)$. While this zero-momentum forward scattering mode provides a thermal mass shift for the heavy quark, it does not contribute to finite-angle scattering. Because the evaluation of transport coefficients, such as the momentum diffusion rate and energy loss, intrinsically incorporates a weighting factor of the momentum transfer squared $|\vec{q}\,|^2$ or energy transfer $\omega$, this term integrates to identically zero. Consequently, this nondynamical constant shift does not alter the heavy-quark kinematic state and is strictly omitted from the interaction kernel. Crucially, the dynamical string component $\tilde{V}_{\rm S}(q, T)$ exhibits a $1/q^4$ asymptotic behavior in the ultraviolet regime, ensuring it smoothly decouples at large momentum transfers whilst dominating the strong interaction in the infrared regime.

\section{Unified Scattering Amplitudes}\label{sec:amplitudes}

The elastic scattering between a heavy quark and a thermalized medium parton (light quarks $q$, antiquarks $\bar{q}$, or gluon $g$) is expressed as
\begin{equation*}
Q(P_1)+i(P_2)\to Q(P_3)+i(P_4),
\end{equation*}
where $P_{1}=(E_{1},\vec{p}_{1})$ and $P_2$ are the four-momenta of the incoming heavy quark ($Q$) and medium parton ($i = q, \bar{q}, g$), respectively, and $P_3$ and $P_4$ are the four-momenta of the outgoing particles.
The four-momentum transfer in each channel is defined as $P^{\mu}_1-P^{\mu}_3=(\omega,\vec{q}\;)=(\omega,\vec{q}_T,q_L)$.

In this work, we adopt an effective field-theoretical construction in which the lattice-constrained in-medium potential is promoted to a dynamical interaction kernel governing heavy-quark scattering. Following earlier transport-based approaches, the analytical Fourier transform of the static potential is interpreted as an effective propagator mediating the real-time interaction between heavy quarks and medium partons~\cite{Riek:2010fk, Xing:2021xwc}. 

Formally, this construction corresponds to a Born-level realization of potential scattering, where the complex gauge field dynamics are encoded into an effective two-body interaction kernel. This approximation is well justified in the heavy-flavor sector by the mass hierarchy $m_Q \gg T$. In the medium rest frame, thermal heavy quarks are predominantly nonrelativistic with velocity $v \sim \sqrt{T/m_Q} \ll 1$. As rigorously demonstrated by Riek and Rapp~\cite{Riek:2010fk}, for heavy-quark elastic scattering in the QGP, this kinematic constraint ensures that the energy transfer is parametrically smaller than the three-momentum transfer ($q_0 \ll |\vec{q}|$ or $\omega \sim v|\vec{q}\,| \ll |\vec{q}\,|$). Therefore, taking the static limit ($q_0 \to 0$) to map the potential to an effective $t$-channel propagator is a well-justified approximation, and the interaction is strictly dominated by spacelike exchange and can be robustly treated as instantaneous. This picture is highly consistent with effective field theory formulations such as potential nonrelativistic QCD (pNRQCD), where the heavy-quark potential emerges as an instantaneous interaction kernel~\cite{Brambilla:1999xf}, as well as with thermodynamic $T$-matrix approaches~\cite{Riek:2010fk}.

A crucial element of this framework is the rigorous separation between the static effective coupling $\tilde{\alpha}_s$ and the dynamical running coupling $\alpha_s(\Lambda_{\text{hard}})$. This distinction stems from their respective domains. The effective coupling $\tilde{\alpha}_s$ is defined within the V scheme to intrinsically characterize the nonperturbative features of the lattice-extracted static potential at soft infrared scales. In contrast, for the dynamical scattering vertices, we employ the standard $\overline{\text{MS}}$ running coupling $\alpha_s(\Lambda_{\text{hard}})$, evaluated at the adaptive hard scale $\Lambda_{\text{hard}} = \max\{2\pi T, \sqrt{-t}\}$, to guarantee asymptotic freedom and prevent unphysical double counting in the perturbative regime.

Guided by this separation, we assign distinct Lorentz structures to the two components of the interaction. The Yukawa term is mediated by a vector interaction vertex, strictly consistent with leading-order perturbative QCD and preserving the correct ultraviolet behavior. In contrast, the long-range string term is modeled using a scalar interaction vertex~\cite{Xing:2021xwc}. While the short-range Coulomb interaction originates from vector gluon exchange, the nonperturbative confining force arises from flux-tube formation, which effectively behaves as an invariant mass generation. This scalar assignment is also strongly supported by phenomenological quarkonium spectroscopy, where a scalar confining potential is requisite to correctly suppress spin-spin interactions and reproduce experimental spin-orbit splittings~\cite{Gromes:1984ma, Lucha:1991vn}. Therefore, a scalar vertex provides the most consistent effective description of the nonperturbative infrared dynamics.

The total scattering amplitude can thus be written as
\begin{equation}
\begin{aligned}\label{eq:Matrix_def}
i\mathcal{M} =& \mathcal{M}_{\rm Y} + \mathcal{M}_{\rm S} \\
=& \bar{u}(P_{3})\gamma^{\mu}u(P_1)\,\tilde{V}_{\rm Y}\,\bar{u}(P_{4})\gamma_{\mu}u(P_2) \\
&+ \bar{u}(P_{3})u(P_1)\,\tilde{V}_{\rm S}\,\bar{u}(P_{4})u(P_2),
\end{aligned}
\end{equation}
where $u(P)$ and $\bar{u}(P)$ denote the fermionic spinors of the external states. 

It is important to emphasize that the present framework is fundamentally constructed at the Born approximation level. While the single-scattering assumption becomes increasingly strained in the strongly coupled regime near $T_c$, its specific role in this study is to establish a unified baseline that continuous incorporates both perturbative and nonperturbative interactions without arbitrary kinematic cutoffs. In this sense, the current approach serves to map the applicability boundaries of static potentials in transport theory and is complementary to more advanced nonperturbative frameworks, such as $T$-matrix resummation, where unitarization is achieved via infinite-order iterations.

Owning to the distinct Lorentz structures of the interaction vertices, the interference term between the Yukawa and string contributions vanishes identically. For heavy-quark scattering with thermal light quarks, this exact cancellation is rigorously guaranteed by chiral symmetry in the massless limit, since the vector interaction is helicity conserving whereas the scalar interaction is helicity flipping~\cite{vanHees:2004gq}. A similar orthogonality holds for gluon scattering under color and Lorentz traces. The squared amplitude therefore cleanly decomposes into independent constituent parts,
\begin{equation}
\overline{|\mathcal{M}|^2} = \overline{|\mathcal{M}|^2}_{\rm Y} + \overline{|\mathcal{M}|^2}_{\rm S}.
\end{equation}

To construct the scattering amplitude from the momentum-space potential, we map the exchange momentum $|\vec{q}\,|^2$ directly to the Mandelstam variable $t$. In the collision center-of-mass frame, the exact identity $t = -|\vec{q}\,|^2$ holds because the energy transfer vanishes identically. For thermal heavy quarks in the medium rest frame ($m_Q \gg T$), the nonrelativistic kinematics ensure that the squared energy transfer $\omega^2$ remains negligible, preserving the robustness of this mapping. Therefore, the interaction kernel is analytically continued into the covariant framework, and the squared matrix elements are evaluated strictly as functions of $t$.

\subsection{Heavy quark and light quark scattering ($Qq \to Qq$)}

To guarantee that the short-range vector interaction exactly reproduces the well-established leading-order perturbative QCD limit for heavy quark and light quark scattering ($Qq \to Qq$), we must properly calibrate the color normalization. Specifically, in the extreme ultraviolet (hard scattering) limit where $|t| \gg M_D^2$, our interaction kernel naturally reproduces the correct pQCD limit. Consistent with implementations in transport models (e.g., Xing $et~al.$~\cite{Xing:2021xwc}), the nonperturbative string contribution decays extremely rapidly as $\sim 1/t^4$ and dynamically decouples. Simultaneously, the Yukawa contribution, governed by the $\overline{\text{MS}}$ running coupling $\alpha_s(\Lambda_{\text{hard}})$, strictly dominates. To ensure that our Yukawa matrix elements analytically converge to the standard leading-order pQCD cross sections with the exact color factor $C_F^2 / (N_c^2 - 1)$ in this $|t| \gg M_D^2$ limit, we explicitly introduce an auxiliary color-average factor $1/(N_c^2-1)$ associated with the effective intermediate field. Because the interaction kernel is treated as a unified whole in our framework, this normalization is uniformly applied to the entire squared amplitude. As a direct consequence, the nonperturbative string contribution also acquires this identical $1/(N_c^2-1)$ scaling factor in the final evaluation.

For the $Qq \to Qq$ scattering process, which proceeds entirely via the $t$ channel, the matrix element squared is given by
\begin{equation}
\label{eq:Amp_Qq_pert}
\overline{|\mathcal{M}_{Qq}|^2}_{\rm Y} = \frac{64\pi^2 \alpha_s^{2}}{9} \frac{\tilde{s}^2 + \tilde{u}^2 + 2m_1^2 t}{(t - M_{\rm D}^2)^2},
\end{equation}
and the string correction is
\begin{equation}
\label{eq:Amp_Qq_string}
\overline{|\mathcal{M}_{Qq}|^2}_{\rm S} = \frac{1}{N_c^2-1} (t^2 - 4m_1^2t) \left[ \frac{8\pi\sigma(t - 5M_{\rm D}^2 )}{(t - M_{\rm D}^2)^3} \right]^2,
\end{equation}
with the kinematic abbreviations $\tilde{s}\equiv s-m_{1}^{2}$ and $\tilde{u}\equiv u-m_{1}^{2}$, where $m_1=m_{Q}$ is the heavy quark mass. The standard Mandelstam relation reads as $\tilde{s}+\tilde{u}+t=0$.

The perturbative part perfectly matches the standard leading-order pQCD result, provided the Debye mass is replaced by the lattice-extracted $M_{\rm D}$. The string part utilizes the explicit Fourier transform of the lattice string potential, multiplied by the appropriate scalar trace factor and color averaging. As previously discussed, in the ultraviolet limit ($|t|\gg M_{\rm D}^2$), the string interaction dynamically decouples, and the full matrix element smoothly reduces to the pure perturbative QCD expectation.

\subsection{Heavy quark and gluon scattering ($Qg \to Qg$)}
The heavy quark and gluon scattering process ($Qg \to Qg$) exhibits a complex structure involving $s$-, $u$-, and $t$-channel exchanges, along with their respective interference terms. Extending our effective potential framework to this process necessitates a rigorous treatment of the distinct interaction scales associated with each channel. The intermediate states in the $s$ and $u$ channels are characterized by high virtuality, representing short-lived excitations that govern hard, short-distance dynamics. Owning to this substantial off shellness, these channels probe distances far smaller than the typical correlation lengths of the medium. Consequently, long-range nonperturbative phenomena, namely Debye screening and string tension, do not have sufficient spacetime to develop, justifying a standard perturbative QCD treatment for the $s$- and $u$-channel amplitudes.

In contrast, the $t$-channel exchange mediates long-range interactions that are highly susceptible to medium polarization and nonperturbative vacuum structures. In the deep infrared regime where $t \to 0$, the medium primarily resolves the total color charge of the interacting parton rather than its internal spin-dependent details. Therefore, we postulate that the nonperturbative string correction applies exclusively to the $t$ channel, acting as an infrared-active scalar interaction. To determine its interaction strength, we invoke the Casimir scaling hypothesis~\cite{Xing:2021xwc}, assuming the effective string potential experienced by a gluon shares the same functional dependence as that for a light quark, but scaled by the color group ratio $C_A/C_F$ to reflect the adjoint representation. To maintain strict theoretical consistency with the quark channel, the auxiliary color-average factor $1/(N_c^2-1)$ is retained to uniformly normalize the interaction kernel. Finally, the perturbative Yukawa contribution is incorporated by utilizing the full leading-order pQCD matrix elements, while rigorously substituting $t \to t - M_{\rm D}^2$ within the $t$-channel denominators to consistently regulate infrared divergences.

The resulting separated amplitude squared is
\begin{equation}
\begin{aligned}\label{eq:Amp_Qg_pert}
\overline{|\mathcal{M}_{Qg}|^2}_{\rm Y} =& 16\pi^2\alpha_s^2 \biggr[ 2\frac{-\tilde{s}\tilde{u}}{(t-M_{\rm D}^2)^{2}} + \frac{4}{9}\frac{-\tilde{s}\tilde{u} + 2m_{1}^{2}(s+m_{1}^{2})}{\tilde{s}^{2}} \\
&+ \frac{4}{9}\frac{-\tilde{s}\tilde{u} + 2m_{1}^{2}(m_{1}^{2}+u)}{\tilde{u}^{2}} + \frac{1}{9}\frac{m_{1}^{2}(4m_{1}^{2}-t)}{-\tilde{s}\tilde{u}} \\
&+ \frac{-\tilde{s}\tilde{u} + m_{1}^{2}(s-u)}{(t-M_{\rm D}^2)\tilde{s}} - \frac{-\tilde{s}\tilde{u} - m_{1}^{2}(s-u)}{-(t-M_{\rm D}^2)\tilde{u}} \biggr],
\end{aligned}
\end{equation}
\begin{equation}
\label{eq:Amp_Qg_string}
\overline{|\mathcal{M}_{Qg}|^2}_{\rm S} = \frac{C_A}{C_F} \frac{1}{N_c^2-1} (t^2 - 4m_1^2t) \left[ \frac{8\pi\sigma(t - 5M_{\rm D}^2 )}{(t - M_{\rm D}^2)^3} \right]^2.
\end{equation}

To maintain thermodynamic consistency, the screening mass $M_{\rm D}$ and the infrared static potentials are evaluated using the running coupling at the thermal scale $\Lambda_{\text{soft}} = 2\pi T$. However, for the hard vertices $\alpha_s$ appearing explicitly in the perturbative amplitudes, we enforce the adaptive renormalization scale $\Lambda_{\text{hard}} = \max \{ 2\pi T, \sqrt{-t} \}$, dynamically regulating the Landau pole and ensuring asymptotic freedom at large momentum transfers.

\section{Heavy-Quark Energy Loss and Transport Coefficients}\label{sec:transport}
Armed with the unified matrix elements, the transport coefficients can be computed continuously over the full phase space. The total interaction rate for a heavy quark scattering off the surrounding thermal medium partons is
\begin{widetext}
\begin{equation}
\begin{aligned}
\Gamma(E_1,T) = \sum_{i=q,g}\Gamma_{i}(E_1,T) = \frac{1}{2E_1}\sum_{i=q,g}\int\frac{d^{3}\vec{p}_2}{(2\pi)^{3}2E_{2}} n_i(E_2,T)\int\frac{d^{3}\vec{p}_3}{(2\pi)^{3}2E_{3}}
\int\frac{d^{3}\vec{p}_4}{(2\pi)^{3}2E_{4}} \overline{|\mathcal{M}^{2}|}_{Qi}(2\pi)^{4}\delta^{(4)}(\Sigma P).
\end{aligned}
\end{equation}

The energy loss per unit path length ($-dE/dz$) and the transverse/longitudinal momentum diffusion coefficients ($\kappa_{T/L}$) are obtained via standard kinetic weighting of the interaction rate~\cite{Lou:2025wmw, Peng:2024zvf}:
\begin{equation}\label{eq:Global_dEdz}
\begin{aligned}
-\frac{dE}{dz} = \int d^3\vec{q}\frac{d\Gamma}{d^3\vec{q}}\frac{\omega}{v_1}
= \frac{1}{256\pi^{3}p_1^{2}} \sum_{i=q,g} \int_{0}^{\infty} dp_{2} E_{2} n_i(E_2,T) \int_{-1}^{1} d(\cos\psi) \int_{t_{min}}^{0} dt \frac{b}{a^{3}} \;
\overline{|\mathcal{M}^{2}_{Qi}|},
\end{aligned}
\end{equation}
\begin{equation}\label{eq:Global_kappaT}
\begin{aligned}
\kappa_{T} = \frac{1}{2}  \int d^{3}\vec{q} \; \frac{d\Gamma}{d^{3}\vec{q}} \; \vec{q}^{\;2}_{T}
= \frac{1}{256\pi^{3}p_1^{3}E_1} \sum_{i=q,g} \int_{0}^{\infty} dp_{2} E_{2} n_{i}(E_{2},T) \int_{-1}^{1} d(\cos\psi) \int_{t_{min}}^{0} dt ~\mathcal{A} \, \overline{|\mathcal{M}^{2}_{Qi}|},
\end{aligned}
\end{equation}
\begin{equation}\label{eq:Global_kappaL}
\begin{aligned}
\kappa_{L} = \int d^{3}\vec{q} \frac{d\Gamma}{d^{3}\vec{q}} \; q^{2}_{L}
= \frac{1}{256\pi^{3}p_1^{3}E_1} \sum_{i=q,g} \int_{0}^{\infty} dp_{2} E_{2} n_{i}(E_{2},T) \int_{-1}^{1} d(\cos\psi) \int_{t_{min}}^{0} dt ~\mathcal{B} \, \overline{|\mathcal{M}^{2}_{Qi}|},
\end{aligned}
\end{equation}
\end{widetext}
with the incident velocity $v_1={p}_1/E_1$ and the abbreviations
\begin{equation}
\begin{aligned}\label{eq:Short_AB}
\mathcal{A} &\equiv \frac{1}{a} \left[ -\frac{m_1^{2}\left(D+2b^{2}\right)}{8p_1^{\;2}a^{4}} + \frac{E_1tb}{2p_1^{\;2}a^{2}} - t\left(1+\frac{t}{4p_1^{\;2}}\right) \right],  \\
\mathcal{B} &\equiv \frac{1}{a} \left[ \frac{E_1^{2}\left(D+2b^{2}\right)}{4a^{4}} -\frac{E_1tb}{a^{2}} + \frac{t^{2}}{2} \right].
\end{aligned}
\end{equation}
The integration boundaries and auxiliary variables in Eqs.~(\ref{eq:Global_dEdz})-(\ref{eq:Short_AB}) are given by
\begin{equation}
\begin{aligned}
&t_{min} = -\frac{\tilde{s}^{2}}{s}, \qquad a = \frac{\tilde{s}}{p_1}, \\
&b =-\frac{2t}{p_1^{2}} \left[ E_1\tilde{s}-E_{2}\left(s+m_1^{2}\right) \right], \\
&c =-\frac{t}{p_1^{2}} \left\{ t\left[ \left(E_1+E_{2}\right)^{2}-s \right] + 4p_1^{2}p_{2}^{2}\sin^{2}\psi \right\}, \\
&D = -t \left(ts + \tilde{s}^{2}\right) \cdot \left( \frac{4E_{2}\sin\psi}{p_1} \right)^{2}.
\end{aligned}
\end{equation}

The local thermal equilibrium distribution functions for bosons (gluons) and fermions (light quarks and antiquarks) are defined as
\begin{equation}\label{eq:ThermalDis_Boson}
n_B(E,T) = \left(e^{E/T}-1\right)^{-1},
\end{equation}
\begin{equation}
\label{eq:ThermalDis_FermionAvg}
n_{F}(E,T) =  \left(e^{E/T} + 1\right)^{-1}.
\end{equation}

Crucially, the integrations over the momentum transfer squared $t$ in Eqs.~(\ref{eq:Global_dEdz})-(\ref{eq:Global_kappaL}) run smoothly from the kinematic minimum $t_{min}$ all the way up to $0$. The infrared divergences are dynamically screened by the lattice-constrained $M_{\rm D}$ residing within our unified matrix elements, completely bypassing the need for an arbitrary soft-hard separation boundary.

Building upon the momentum-space diffusion coefficients evaluated above, we can further extract the jet transport coefficient $\hat{q}$. The parameter $\hat{q}$ characterizes the average transverse momentum squared accumulated by a hard parton per unit path length as it propagates through the thermal medium. It serves as a fundamental quantity in describing the jet quenching phenomenon in ultrarelativistic heavy-ion collisions. This transport coefficient is directly proportional to the transverse momentum diffusion rate $\kappa_T$ and inversely proportional to the heavy quark velocity $v$, expressed as
\begin{equation}\label{eq:qhat}
\hat{q} = \frac{2\kappa_T}{v}.
\end{equation}

Furthermore, in the static limit where the heavy quark velocity approaches zero ($v \to 0$), the momentum diffusion process becomes isotropic. In this scenario, the transverse and longitudinal diffusion coefficients converge to a uniform value, denoted as $\kappa_T \approx \kappa_L \equiv \kappa$. This zero-momentum diffusion coefficient $\kappa$ is intimately connected to the macroscopic spatial diffusion coefficient $D_s$ through the fluctuation-dissipation theorem. To facilitate direct comparisons with lattice QCD calculations and phenomenological models, it is conventional to express this relationship in a dimensionless form:
\begin{equation}\label{eq:Ds}
2\pi T D_s = \frac{4\pi}{\kappa/T^3}.
\end{equation}
The quantity $2\pi T D_s$ governs the spatial relaxation and the Brownian motion of heavy quarks within the quark-gluon plasma~\cite{Moore:2004tg}. It acts as a critical indicator of the medium coupling strength, where a smaller value of $2\pi T D_s$ corresponds to a stronger interaction between the heavy quarks and the thermal bath.

\section{Results and Discussion}\label{sec:results}
\begin{figure}[!htbp]
\begin{center}
\setlength{\abovecaptionskip}{-0.1mm}
\includegraphics[width=.47\textwidth]{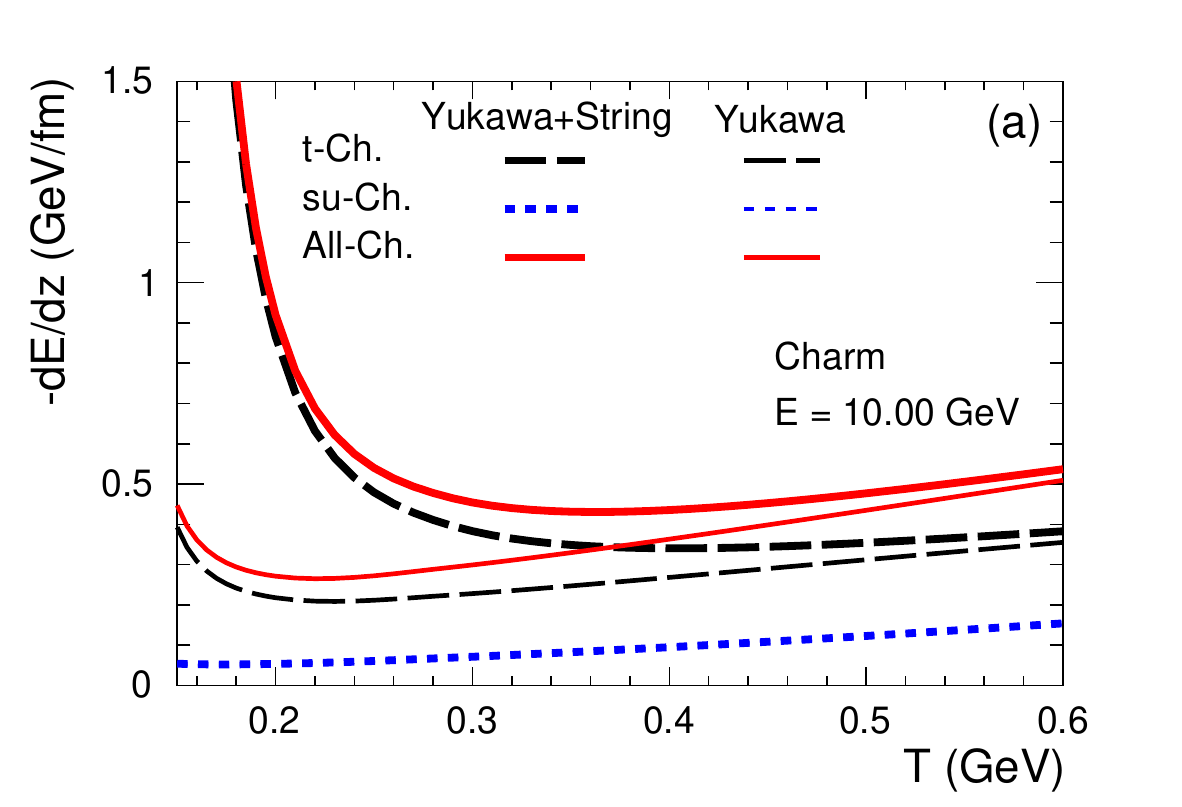}
\includegraphics[width=.47\textwidth]{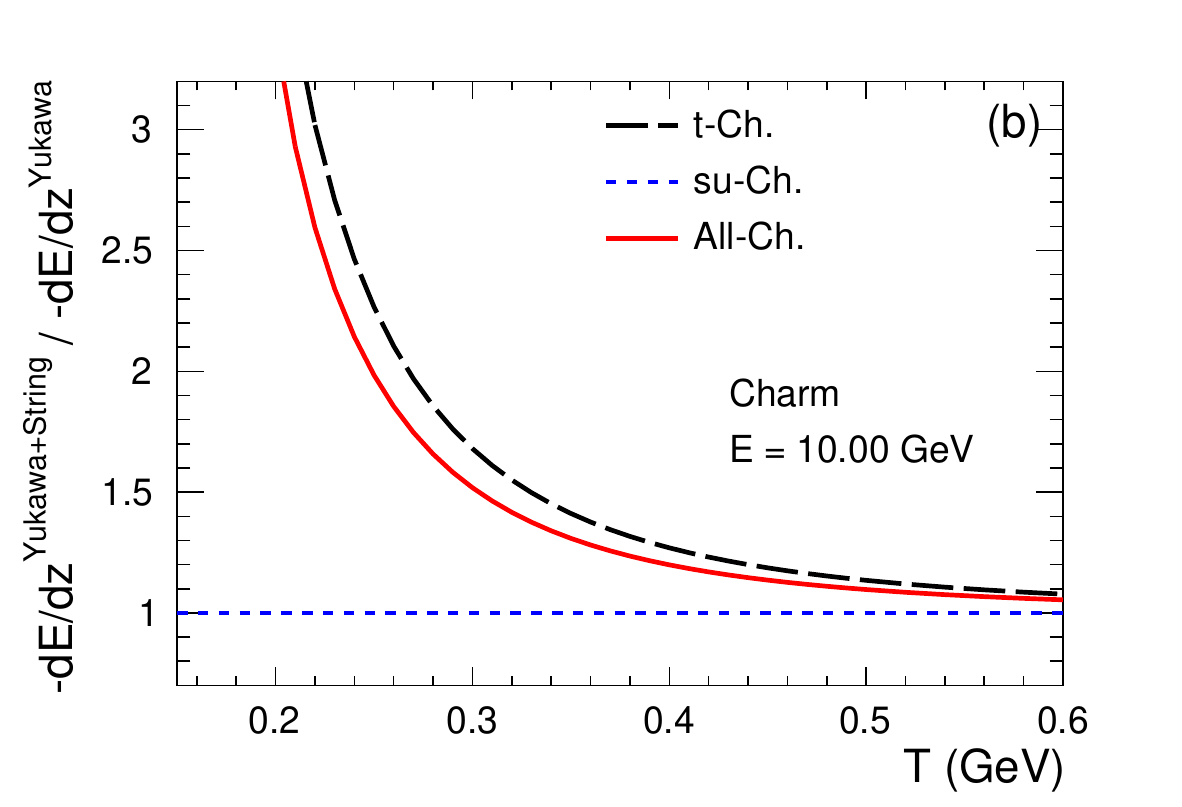}
\caption{(a) The spatial energy loss $-dE/dz$ of a charm quark with a fixed incident energy $E=10~{\rm GeV}$ as a function of temperature. The results are calculated using the full effective potential (labeled as Yukawa+String, thick curves) and the leading-order perturbative baseline (labeled as Yukawa, thin curves). Contributions from various partonic channels are displayed separately using different line styles. (b) The relative ratio of the energy loss between the full approach and the perturbative baseline.}
\label{fig:Charm_dEdz_vsT}
\end{center}
\end{figure}
Figure~\ref{fig:Charm_dEdz_vsT} illustrates the temperature dependence of the spatial energy loss $-dE/dz$ for a charm quark traversing the thermal medium with a fixed incident energy of $E=10~{\rm GeV}$. To isolate the impact of nonperturbative dynamics, we explicitly compare the full model evaluation, which incorporates both the Yukawa and string interactions, against a pure leading-order perturbative Yukawa baseline. The bottom panel, displaying the relative ratio between these two scenarios, clearly reveals that the deviation is highly localized in the low-temperature regime. Specifically, the inclusion of the string term significantly enhances the energy loss at temperatures $T \lesssim 1.7T_c$ (approximately $0.3~{\rm GeV}$), whereas the ratio rapidly approaches unity as the temperature further increases.

As illustrated by the separate partonic channel contributions in the bottom panel, the $s$- and $u$-channel amplitudes remain identical and purely perturbative in both scenarios. Therefore, the observed nonperturbative enhancement is driven exclusively by the $t$-channel exchange. This localized enhancement is a direct consequence of the strongly coupled nature of the quark-gluon plasma near the phase transition. In the vicinity of $T_c$, the medium is densely populated with nonperturbative vacuum remnants, allowing the long-range spatial string tension to strongly scatter the propagating heavy quark via the $t$ channel. As the temperature rises, the Debye screening mass increases substantially, dynamically melting the confining flux tubes. Consequently, the long-range string interaction is effectively screened out, and the thermal scattering asymptotically reduces to the pure short-range perturbative exchange. 

\begin{figure}[!htbp]
\begin{center}
\setlength{\abovecaptionskip}{-0.1mm}
\includegraphics[width=.47\textwidth]{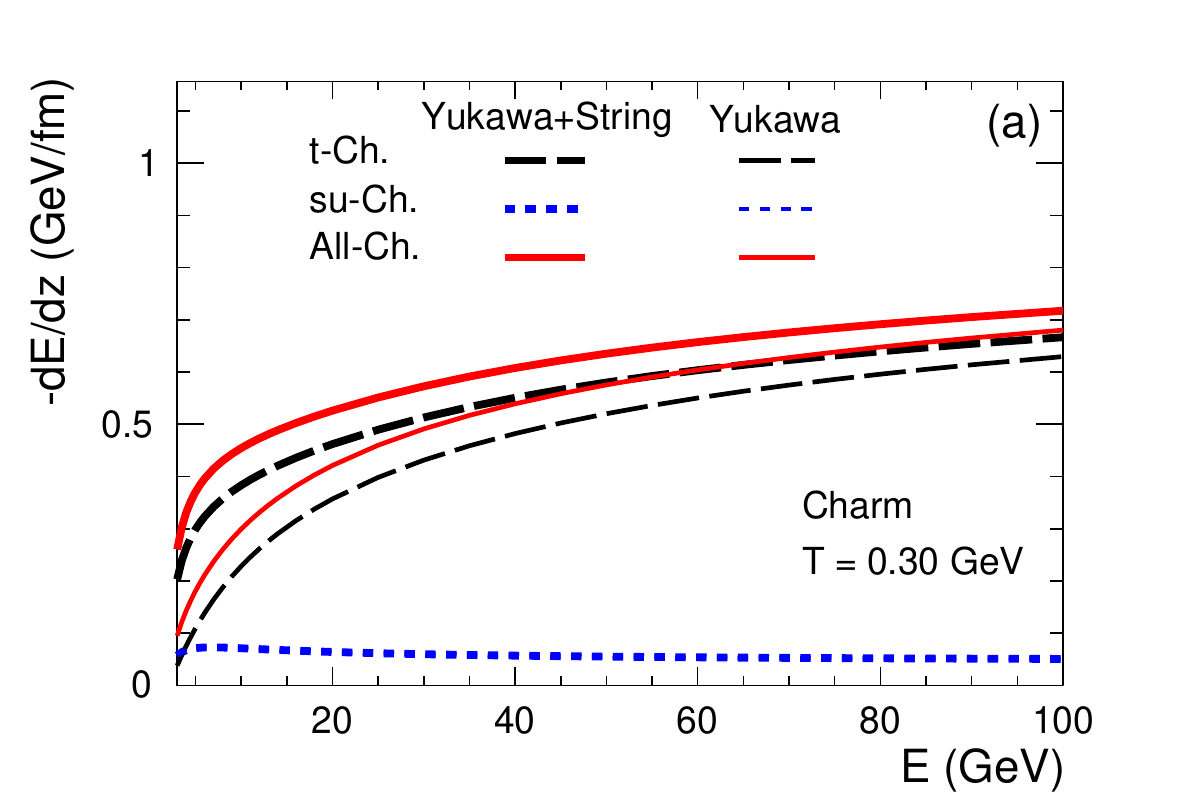}
\includegraphics[width=.47\textwidth]{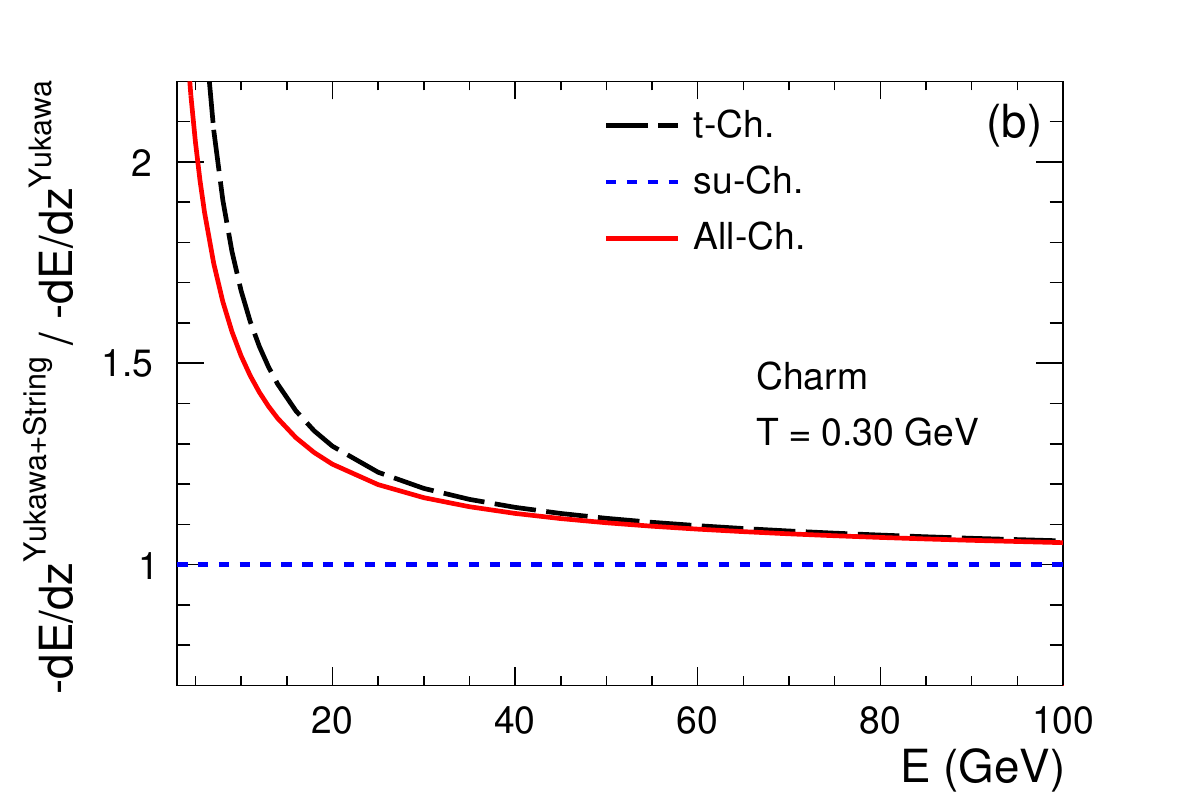}
\caption{Same as Fig.~\ref{fig:Charm_dEdz_vsT}, but presented as a function of the heavy-quark incident energy at a fixed medium temperature of $T=0.3~{\rm GeV}$.}
\label{fig:Charm_dEdz_vsE}
\end{center}
\end{figure}
To further dissect the kinematic behavior of this nonperturbative enhancement, Fig.~\ref{fig:Charm_dEdz_vsE} presents the energy loss as a function of the heavy-quark incident energy at a fixed temperature $T=0.3~{\rm GeV}$, a region where the string effect was previously shown to be substantial. The comparison shows that the discrepancy between the full model and the Yukawa baseline is most pronounced in the low-energy domain. As the incident energy of the heavy quark increases, the relative deviation gradually diminishes, indicating that highly energetic quarks are less sensitive to the infrared string tension.

This behavior naturally arises from the distinct momentum transfer scales associated with the scattering kinematics. For a low-energy heavy quark, the typical momentum transfer during collisions is heavily weighted toward the soft infrared regime, where the nonperturbative string interaction dominates the cross section. In contrast, an energetic heavy quark undergoes collisions with significantly larger momentum transfers, effectively probing the short-distance structure of the interaction kernel. In this ultraviolet limit, the string potential decays asymptotically as $1/q^4$, which is heavily suppressed compared to the $1/q^2$ Coulomb-like Yukawa interaction. Therefore, high-energy heavy quarks inherently bypass the long-range confining interactions and primarily experience perturbative energy loss.

\begin{figure}
\begin{center}
\setlength{\abovecaptionskip}{-0.1mm}
\includegraphics[width=.47\textwidth]{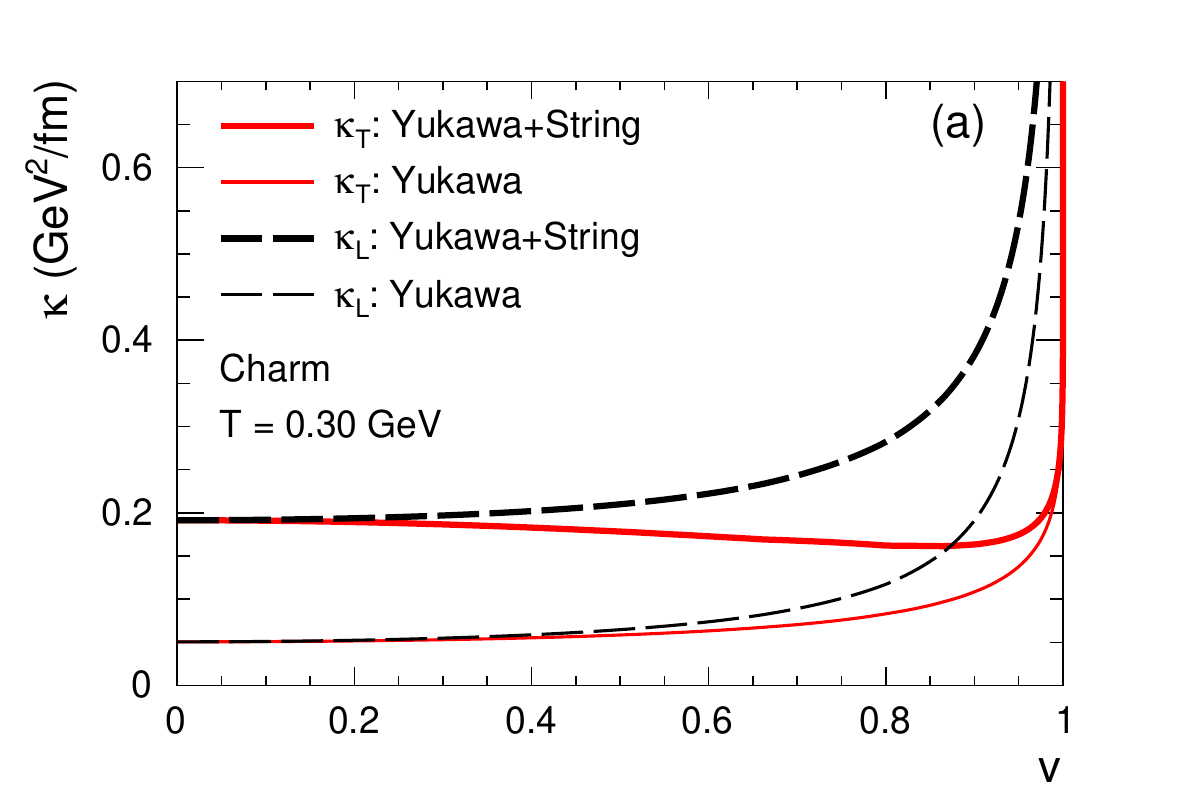}
\includegraphics[width=.47\textwidth]{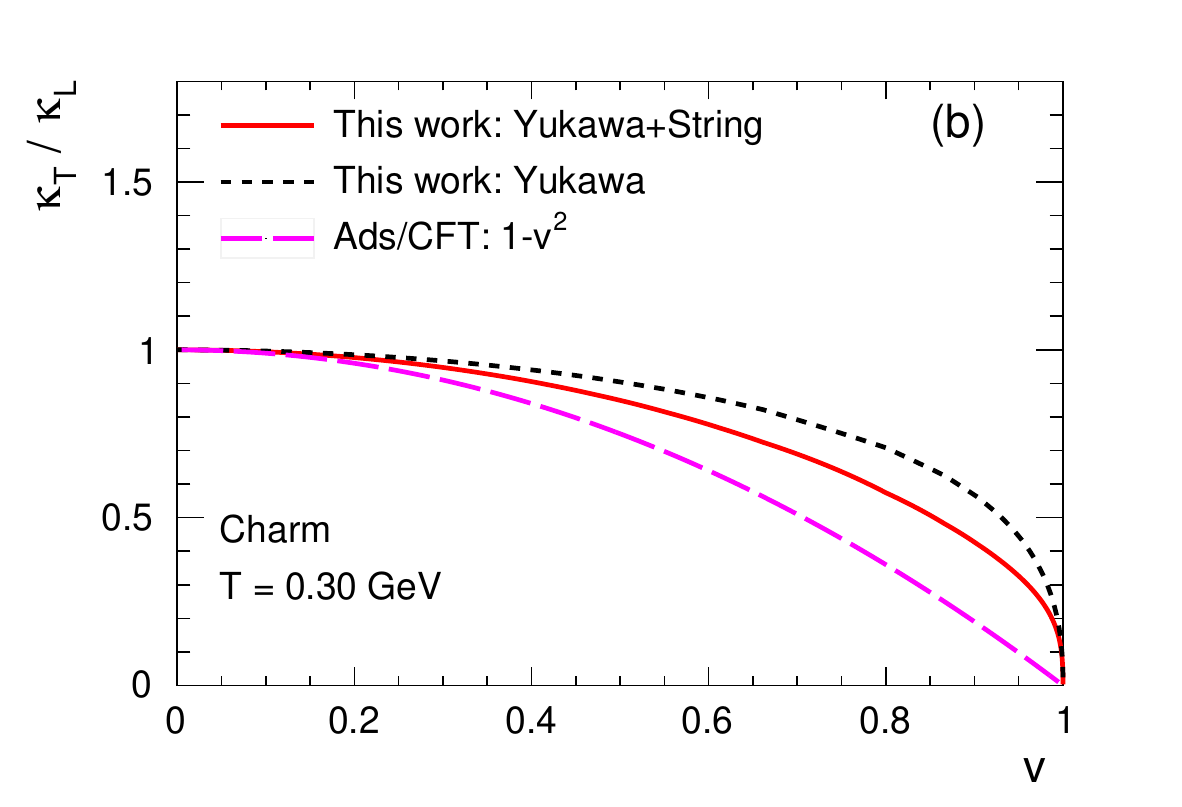}
\caption{(a) The transverse ($\kappa_T$) and longitudinal ($\kappa_L$) momentum diffusion coefficients as functions of the heavy-quark velocity $v$ at a fixed medium temperature of $T=0.3~{\rm GeV}$. The thick curves represent the full model evaluation (Yukawa+String), and the thin curves denote the perturbative baseline (Yukawa). (b) The ratio of the transverse to longitudinal diffusion coefficients $\kappa_T/\kappa_L$ versus the heavy-quark velocity $v$. The model calculations are compared to the prediction from the AdS/CFT correspondence~\cite{Gubser:2006bz, Gubser:2006nz, Casalderrey-Solana:2007ahi} (long dashed pink curve).}
\label{fig:Charm_KappaT_KappaL_vsV}
\end{center}
\end{figure}
To explicitly quantify the momentum-space anisotropy generated by the lattice-constrained interactions, Fig.~\ref{fig:Charm_KappaT_KappaL_vsV} presents the transverse ($\kappa_T$) and longitudinal ($\kappa_L$) momentum diffusion coefficients, along with their ratio, as functions of the heavy-quark velocity $v$ at $T=0.3~{\rm GeV}$. In the near-static limit ($v\to 0$), the ratio strictly approaches unity for all scenarios. This isotropy is a direct consequence of rotational symmetry in the medium rest frame, confirming that longitudinal and transverse fluctuations are indistinguishable for a static probe. As the heavy-quark velocity increases, $\kappa_T/\kappa_L$ monotonically decreases toward zero. This suppression reflects that kinematic constraints and the redistribution of energy-momentum transfers make longitudinal broadening increasingly dominant relative to transverse broadening for highly relativistic probes.

The comparison between the full evaluation (Yukawa+String) and the perturbative baseline (Yukawa) in Fig.~\ref{fig:Charm_KappaT_KappaL_vsV}(b) reveals how the nonperturbative string tension distinctively affects the broadening process. At finite velocities, the ratio $\kappa_T/\kappa_L$ derived from the full model is systematically suppressed compared to the pure Yukawa baseline. This pronounced anisotropy originates from the distinct Lorentz structure and the infrared dependence of the confining interaction. Modeled as a scalar exchange, the string term incorporates a strong $1/t^2$ factor in the squared amplitude, which enforces an extreme forward-peaking angular distribution. This long-range nature ensures that the string mechanism predominantly drives soft scatterings with infinitesimal transverse momentum transfers. Consequently, while the string tension drastically amplifies the overall magnitude of the momentum diffusion at low velocities [Fig.~\ref{fig:Charm_KappaT_KappaL_vsV}(a)], its forward-peaking feature shifts the scattering phase space to contribute more efficiently to the longitudinal broadening than to the transverse broadening. This dynamic angular redistribution leads to a stronger fractional reduction in $\kappa_T$ relative to $\kappa_L$, thereby driving the ratio closer to the strongly coupled expectation from the AdS/CFT correspondence~\cite{Gubser:2006bz, Gubser:2006nz, Casalderrey-Solana:2007ahi}.

\begin{figure}
\begin{center}
\setlength{\abovecaptionskip}{-0.1mm}
\includegraphics[width=.47\textwidth]{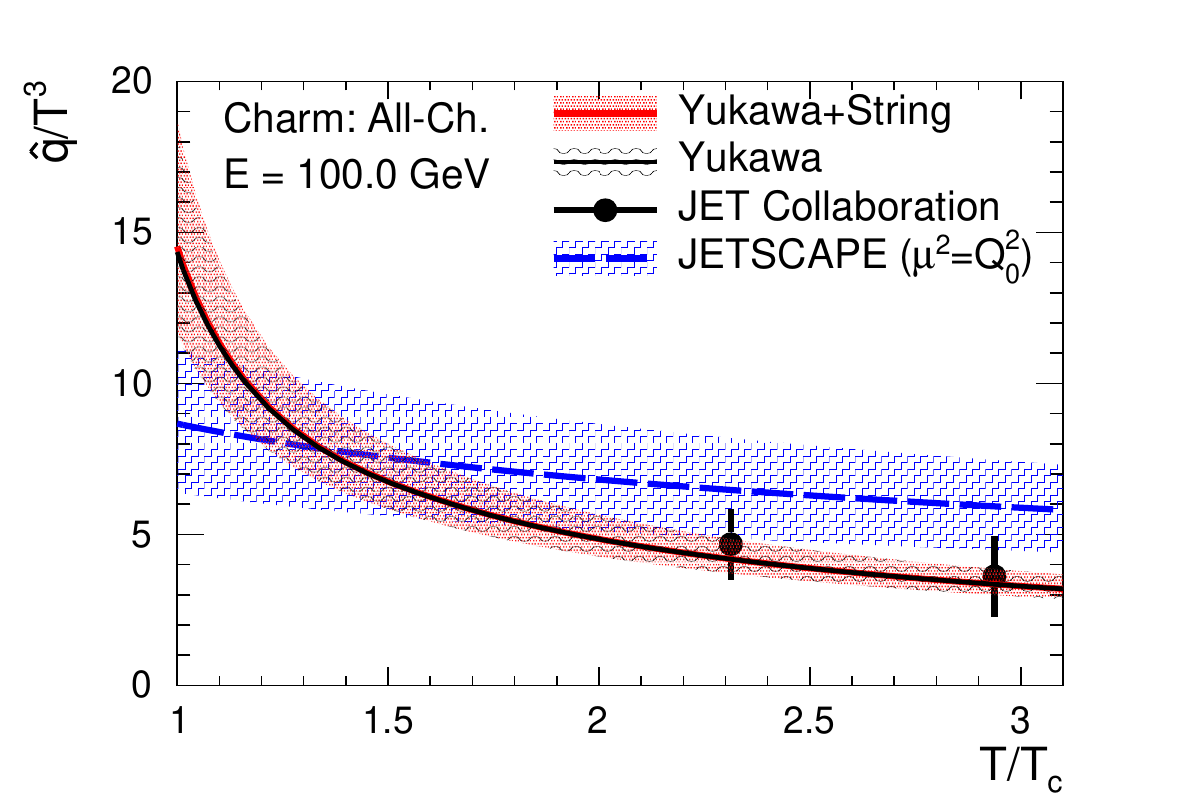}
\caption{The dimensionless transport coefficient $\hat{q}/T^{3}$ as a function of the scaled temperature $T/T_c$ for a charm quark. The static potential model calculations, including both the full approach (shadowed red band) and the leading-order Yukawa baseline (shadowed black band), are performed at a high incident energy of $E=100~{\rm GeV}$. The central curves denote the results evaluated with the central value of the effective screening mass $M_{\rm D}$ while the shaded bands indicate the uncertainties arising from the variation of $M_{\rm D}$. Phenomenological extractions and theoretical predictions are presented for comparison, including the JET Collaboration at $E=10~{\rm GeV}$ (black circles~\cite{JET:2013cls}) and the JETSCAPE Collaboration at $E=100~{\rm GeV}$ (shadowed blue band~\cite{Ehlers:2024miy}).}
\label{fig:Charm_Qhat_vsT}
\end{center}
\end{figure}
Figure~\ref{fig:Charm_Qhat_vsT} presents the scaled momentum broadening rate $\hat{q}/T^3$ for a high-energy charm quark fixed at $E=100~{\rm GeV}$. The theoretical uncertainty of our model, represented by the shaded bands, is rigorously determined by varying the effective screening mass $M_{\rm D}$ within its established theoretical bounds. In striking contrast to the low-energy scenarios discussed previously, comparing the full calculation with the pure Yukawa baseline at $E=100~{\rm GeV}$ reveals almost no visible deviation. The full potential result is nearly identical to the Yukawa-alone baseline. This behavior is perfectly consistent with the kinematic energy dependence established in Fig.~\ref{fig:Charm_dEdz_vsE}. Highly energetic heavy quarks predominantly undergo hard scatterings characterized by large momentum transfers. In this ultraviolet regime, the interaction is deeply penetrating and bypasses the long-range nonperturbative confining force, leaving the perturbative Coulomb-like short-range interaction to completely dominate the momentum diffusion.

To place our high-energy predictions in a broader context, we compare our results with phenomenological extractions and other theoretical frameworks. Despite the purely perturbative nature of the interaction at this high incident energy, our model exhibits a distinct temperature dependence characterized by a prominent peak near the critical temperature followed by a monotonic decrease. This calculation provides a favorable quantitative agreement with the phenomenological extractions from the JET Collaboration~\cite{JET:2013cls}, which were evaluated at $E=10~{\rm GeV}$. More importantly, we compare our results directly with the dynamic extractions from the JETSCAPE Collaboration performed at the identical heavy-quark energy of $E=100~{\rm GeV}$ with $\mu^2=Q_0^2$. Our static potential model successfully captures the qualitative temperature evolution exhibited by the JETSCAPE results.

A detailed comparison with the JETSCAPE results reveals crucial dynamic nuances regarding the interaction kernel. In the immediate vicinity of $T_c$, our model yields a slightly larger $\hat{q}/T^3$ value than JETSCAPE. This enhancement suggests that the lattice-constrained static potential intrinsically encapsulates a stronger residual nonperturbative interaction in the crossover region than the dynamic perturbative kernels typically employed in high-energy event generators. Conversely, at higher temperatures ($T \gtrsim 1.5 T_c$), our model underestimates the JETSCAPE extractions. This systematic transition provides another rigorous consistency check for our framework. As established in the spatial diffusion analysis, the static potential paradigm strictly governs elastic collisions. At high temperatures and high incident energies, dynamic inelastic processes, specifically medium-induced soft gluon radiation, become essential mechanisms for generating transverse momentum kicks. The explicit omission of these radiative channels in our static potential framework naturally leads to an insufficient momentum exchange rate in the high-temperature weakly coupled plasma.

\begin{figure}[!htbp]
\begin{center}
\setlength{\abovecaptionskip}{-0.1mm}
\includegraphics[width=.47\textwidth]{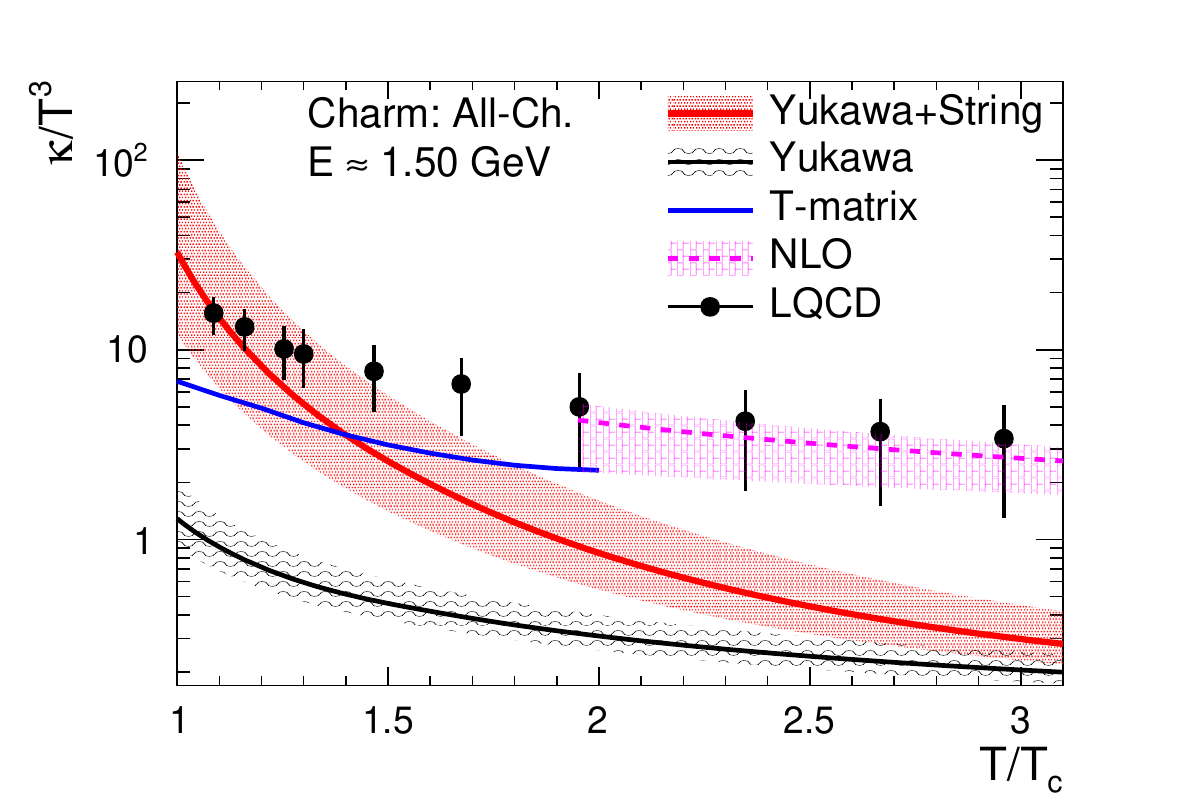}
\caption{The dimensionless heavy-quark momentum diffusion coefficient $\kappa/T^3$ as a function of the scaled temperature $T/T_c$. The calculations are performed for a charm quark with an initial energy of $E\approx1.5$ GeV, corresponding to the static limit ($v \to 0$) where the momentum diffusion anisotropy vanishes ($\kappa_T \approx \kappa_L = \kappa$). The central curves denote the results evaluated with the central value of the effective screening mass $M_{\rm D}$ while the shaded bands indicate the uncertainties arising from the variation of $M_{\rm D}$. The complete model evaluation, incorporating both Yukawa and string interactions, is compared with the pure Yukawa baseline. Lattice QCD data~\cite{HotQCD:2025fbd} and predictions from the $T$-matrix approach~\cite{Riek:2010fk, He:2012df} are presented for comparison. The next-to-leading order (NLO) pQCD calculation~\cite{Caron-Huot:2007rwy} is displayed as a pink band for temperatures above $2T_c$, with the central value corresponding to the renormalization scale $\mu=2\pi T$ and the boundaries indicating the scale variation from $\mu=1.5\pi T$ to $\mu=2.5\pi T$.}
\label{fig:Charm_kappaOverT3}
\end{center}
\end{figure}

Figure~\ref{fig:Charm_kappaOverT3} shows the temperature dependence of the dimensionless momentum diffusion coefficient $\kappa/T^3$ for a charm quark at $E \approx 1.5$ GeV. In this low-velocity limit ($v \to 0$), the momentum diffusion anisotropy vanishes, yielding $\kappa_T \approx \kappa_L = \kappa$. The central curves denote the results evaluated with the central value of the effective screening mass $M_{\rm D}$, with the shaded bands indicating the uncertainties generated by varying the mass within its established bounds. We first compare the complete model evaluation, which incorporates both the Yukawa and string interactions, with the pure Yukawa baseline. The pure Yukawa contribution remains significantly smaller than the full result, particularly near the critical temperature $T_c$.

A quantitative comparison with the latest HotQCD 2025 lattice data~\cite{HotQCD:2025fbd} highlights the indispensable role of this nonperturbative string term. In the critical region $T_c<T\lesssim 1.5T_c$, the pure Yukawa baseline accounts for less than $30\%$ of the total momentum diffusion coefficient. Confronted with the lattice QCD results, the inclusion of the string interaction exactly captures this missing strength, yielding consistent results within theoretical uncertainties in the vicinity of $T_c$. This remarkable agreement demonstrates that the spatial string tension successfully encapsulates the strong nonperturbative dynamics, confirming the survival of confining correlations across the QCD phase transition. The prediction from the $T$-matrix approach~\cite{Riek:2010fk, He:2012df} is also presented, which exhibits good agreement with the lattice data across the entire considered temperature range.

To assess the high-temperature weak-coupling limit, Fig.~\ref{fig:Charm_kappaOverT3} further incorporates the NLO pQCD predictions derived by Caron-Huot and Moore~\cite{Caron-Huot:2007rwy}. This NLO evaluation is presented for temperatures above $2T_c$. In our framework, the interaction kernel is fundamentally constructed at the Born approximation [leading order (LO) ]level, augmented by the nonperturbative string interaction. As the temperature increases into the deep deconfined phase, the long-range string tension dynamically melts, and our full model naturally degenerates to the pure Yukawa baseline. As expected, this asymptotic Yukawa baseline is located significantly below the NLO pQCD band. The NLO calculations incorporate large positive $\mathcal{O}(g)$ corrections from higher-order perturbative effects, such as overlapping scatterings, and demonstrate a good agreement with the lattice QCD data within theoretical uncertainties in this high-temperature regime. The visible gap between our asymptotic LO baseline and the NLO band accurately delineates the magnitude of these higher-order perturbative corrections. Conversely, in the vicinity of $T_c$, our full model incorporating the string tension provides a massive nonperturbative enhancement and overtakes the perturbative expectations. This hierarchy elegantly validates that the extreme opacity of the QGP near the critical temperature is predominantly driven by nonperturbative infrared dynamics, whereas high-temperature momentum diffusion necessitates the inclusion of higher-order perturbative scatterings.

The systematic deviation of our model at higher temperatures ($T \gtrsim 1.5T_c$) provides critical insight into the intrinsic applicability boundary of the static potential paradigm. The current framework reliably encapsulates the nonperturbative infrared enhancement driven by elastic scattering. Nevertheless, as the temperature increases and the running coupling weakens, dynamic inelastic processes, specifically medium-induced soft gluon radiation, emerge as the dominant mechanism for momentum broadening. Generating real radiative emission fundamentally requires dynamic energy transfer ($\omega > 0$). This kinematic requirement falls strictly outside the domain of any instantaneous static potential approximation, which inherently relies on the $\omega \approx 0$ limit. Consequently, the observed discrepancy precisely demarcates the temperature threshold where dynamical and inelastic transport phenomena must be resummed beyond the single-scattering static limit.

\begin{figure}[!htbp]
\begin{center}
\setlength{\abovecaptionskip}{-0.1mm}
\includegraphics[width=.47\textwidth]{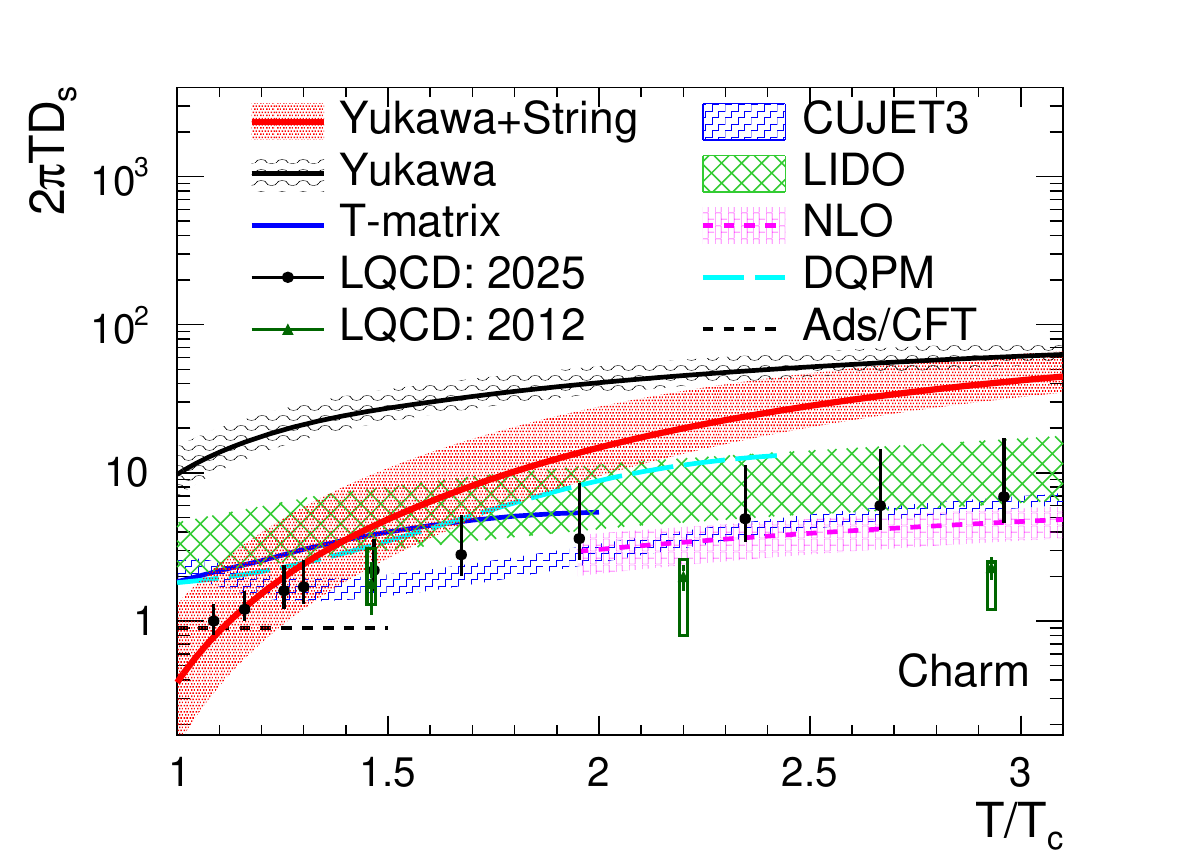}
\caption{The dimensionless spatial diffusion coefficient $2\pi TD_s$ for a charm quark as a function of the scaled temperature $T/T_c$, extracted via Eq.~(\ref{eq:Ds}). The central curves denote the results evaluated with the central value of the effective screening mass $M_{\rm D}$ while the shaded bands indicate the uncertainties arising from the variation of $M_{\rm D}$. The static potential model results are compared with lattice QCD data~\cite{HotQCD:2025fbd, Ding:2012sp} and dynamic phenomenological frameworks, including the $T$ matrix~\cite{Riek:2010fk, He:2012df}, CUJET3~\cite{Shi:2018izg, Shi:2018lsf, Xu:2015bbz}, and LIDO~\cite{Ke:2018tsh}. The prediction from the dynamical quasiparticle model (DQPM)~\cite{Sambataro:2025obe, Sambataro:2024mkr} is shown as the long dashed cyan curve. The strongly coupled AdS/CFT prediction~\cite{Casalderrey-Solana:2006fio} is represented by the dashed black curve.}
\label{fig:Charm_2PiTDs}
\end{center}
\end{figure}

In Fig.~\ref{fig:Charm_2PiTDs}, we present the corresponding temperature dependence of the dimensionless spatial diffusion coefficient $2\pi TD_s$ of charm quark, derived directly from the momentum diffusion coefficient via Eq.~(\ref{eq:Ds}). Specifically, our model predicts $2\pi TD_s$ ranging from 0.5 to 1.7 in the immediate vicinity of $T_c$, exhibiting a remarkable quantitative agreement with the state-of-the-art lattice QCD results from the HotQCD Collaboration (2025)~\cite{HotQCD:2025fbd}, which report $2\pi TD_s=1.0^{+0.3}_{-0.2}$ at $T\approx1.09T_c$. This convergence is highly nontrivial. While the perturbative Yukawa interaction alone would yield a significantly larger diffusion coefficient, the inclusion of the lattice-constrained string tension provides the necessary nonperturbative enhancement to capture the extreme opacity of the QCD medium near the phase transition. Consistent with the $\kappa/T^3$ behavior, our model aligns well with the lattice QCD data~\cite{HotQCD:2025fbd, Ding:2012sp} within theoretical uncertainties near $T_c$. However, in the high-temperature regime, our framework significantly overestimates the spatial diffusion.

This overestimation is a rigorous consequence of the interaction kernel limitations discussed above. Because $D_s$ is inversely proportional to the momentum broadening rate $\kappa$, the explicit omission of radiative energy loss channels directly results in an undersuppressed spatial diffusion. In contrast, advanced phenomenological approaches such as CUJET3~\cite{Shi:2018izg, Shi:2018lsf, Xu:2015bbz} and LIDO~\cite{Ke:2018tsh} explicitly incorporate medium-induced radiative processes, while the $T$-matrix framework~\cite{Riek:2010fk, He:2012df} utilizes a dynamically resummed interaction kernel. Their success in reproducing the high-temperature lattice data reinforces our central theoretical assertion: The instantaneous static interaction strongly governs the phase transition region but must be supplemented by dynamic radiation mechanisms to correctly characterize the weakly coupled quark-gluon plasma.

Furthermore, to provide a holistic view of the strongly coupled QGP dynamics, we comprehensively compare our results with the DQPM~\cite{Sambataro:2025obe, Sambataro:2024mkr} and predictions from the strongly coupled AdS/CFT correspondence~\cite{Casalderrey-Solana:2006fio}. The AdS/CFT framework yields a nearly constant value of $2\pi T D_s \approx 0.9$, which coincides with the lattice QCD extractions within error bars near the critical temperature. Remarkably, our full potential model generates a diffusion coefficient that closely approaches this holographic strong-coupling limit in the crossover region. This convergence powerfully reinforces the consensus that the QGP is inherently strongly coupled near the phase transition. On the other hand, the DQPM elegantly captures the nonperturbative QGP thermodynamics by fitting massive quasiparticle dispersions to the lattice equation of state. When compared to the lattice data, the DQPM yields a higher spatial diffusion coefficient in the lower temperature region $T \lesssim 1.26 T_c$. However, in the intermediate temperature range $1.26 T_c \lesssim T \lesssim 2.43 T_c$, the DQPM results align well with the lattice extractions within the established uncertainty boundaries. This distinct temperature evolution between the lattice-constrained potential approach and the quasiparticle picture highlights the complementary nature of these phenomenological models in decoding the complex medium dynamics.

\begin{figure}[!htbp]
\begin{center}
\setlength{\abovecaptionskip}{-0.1mm}
\includegraphics[width=.47\textwidth]{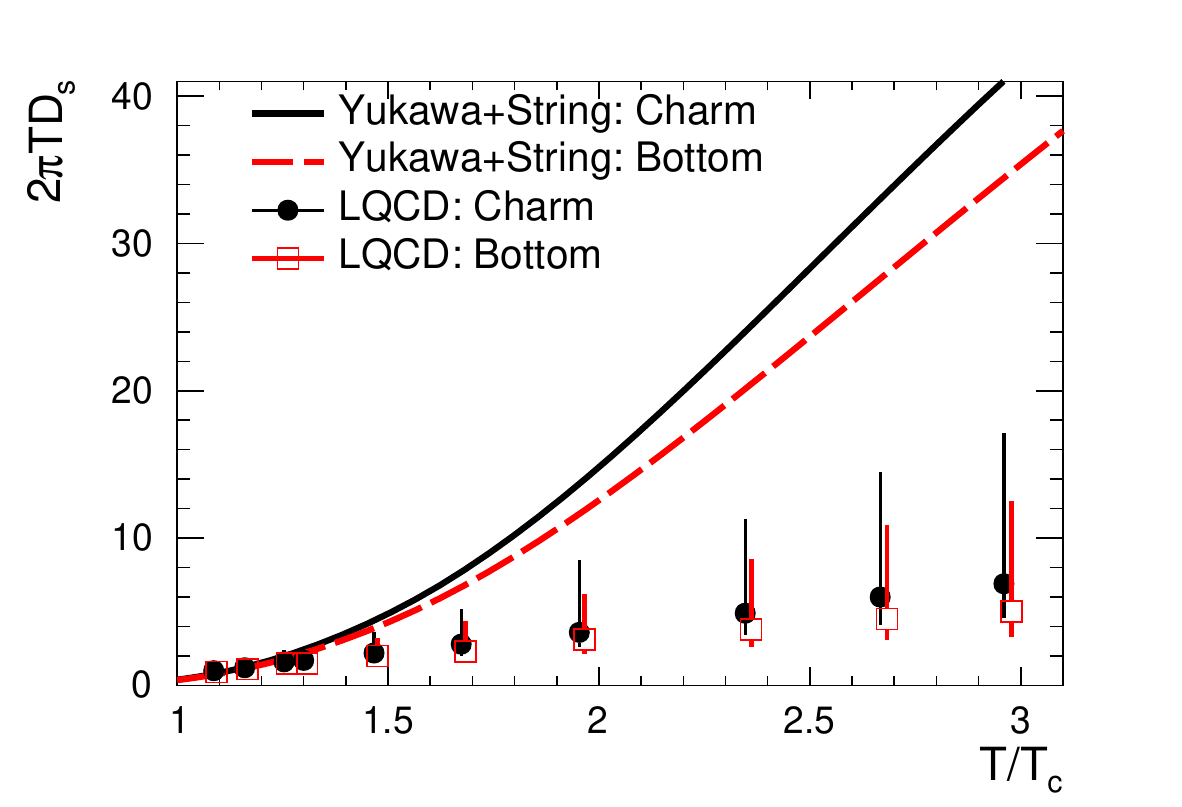}
\caption{Comparison of the spatial diffusion coefficient $2\pi T D_s$ between charm and bottom quarks as a function of the scaled temperature $T/T_c$. The solid black curve and the long dashed red curve represent the central values of our model evaluations for charm and bottom quarks, respectively. To maintain visual clarity, the theoretical uncertainty bands generated by varying the screening mass are omitted; the uncertainty magnitude for the bottom quark is comparable to that of the charm quark shown in Fig.~\ref{fig:Charm_2PiTDs}. Lattice QCD data for charm (black circles) and bottom quarks (open red squares) are included for comparison, with the bottom data slightly shifted horizontally for visibility.}
\label{fig:Charm_Bottom_Ds}
\end{center}
\end{figure}
To rigorously investigate the mass hierarchy of heavy-flavor transport, we extend our evaluation to the bottom quark sector. Figure~\ref{fig:Charm_Bottom_Ds} displays a comparative analysis of the spatial diffusion coefficients for charm and bottom quarks. For visual clarity, only the central values of the theoretical evaluations are presented. The systematic uncertainties for the bottom quark, originating from the variation of the screening mass, are found to be of similar magnitude to those of the charm quark and do not alter the observed mass hierarchy trend.

In the immediate vicinity of the critical temperature, our model predicts a remarkable near degeneracy between the two flavors, yielding a bottom-to-charm $D_s$ ratio close to unity. This behavior indicates that the strong nonperturbative string interaction provides a universal scattering strength that largely overwhelms the kinematic mass differences in the strongly coupled crossover region. However, as the temperature increases into the deconfined phase ($T_c<T<3T_c$), a noticeable mass dependence gradually emerges. Our calculation shows that the bottom-to-charm $D_s$ ratio steadily decreases from approximately 0.93 to 0.83. This dynamic separation perfectly mirrors the trend observed in the corresponding lattice QCD extractions, where the ratio drops from 0.94 to 0.72. This quantitative consistency confirms that our unified interaction kernel seamlessly captures the transition from a mass-independent strongly coupled regime near $T_c$ to a kinematics-dominated weakly coupled regime at higher temperatures.

\section{Conclusion and Outlook}\label{sec:summary}
In this work, we have reformulated the theoretical description of heavy-quark transport in the strongly coupled QCD medium by upgrading the traditional soft-hard factorization approach into a unified, boundary-free framework. By extracting the effective screening mass and long-range string tension from lattice QCD static potentials and analytically Fourier transforming them into momentum space, we derived a continuous interaction kernel capable of governing collision dynamics across the entire phase space. In particular, our framework successfully integrates nonperturbative enhancements, especially the confining string interactions vital in the near-$T_c$ regime, while automatically preserving the rigorous pQCD asymptotic behavior at high momentum transfers via an adaptive running coupling scheme. The elimination of the artificial separation scale $t^\ast$ significantly reduces model dependencies compared to traditional approaches.

Beyond yielding favorable numerical agreement with lattice data, our findings suggest that heavy-quark transport coefficients serve as a high-fidelity probe of the evolving degrees of freedom within the QCD crossover. The peak structure observed in the momentum broadening rate $\hat{q}/T^3$ reflects the survival of nonperturbative confining correlations, which dominate the scattering dynamics even above the pseudocritical temperature. Ultimately, this framework establishes a robust link between static lattice observables and real-time transport dynamics, providing a fundamental baseline for high-precision heavy-flavor phenomenology. 

While the current methodology demonstrates substantial progress in bridging static lattice observations with real-time dynamical transport, we note that the present framework is predominantly driven by chromoelectric interactions. Furthermore, as clearly demarcated by the high-temperature diffusion coefficients, the instantaneous static potential paradigm naturally limits its validity to the regime dominated by elastic scattering. The precise inclusion of dynamical chromomagnetic screening effects and the rigorous resummation of inelastic radiative processes remain important objectives for future theoretical investigations.

Regarding phenomenological validation, our immediate next step is to implement the heavy-quark momentum diffusion coefficients derived from this unified potential model into our previously developed Langevin-transport with gluon radiation framework~\cite{Li:2019lex, Li:2019wri, Li:2018izm, Li:2018jba, Li:2020umn}. By embedding these rigorous microscopic transport coefficients into a macroscopic dynamical evolution model, we aim to systematically evaluate existing experimental measurements of heavy-flavor observables, such as the nuclear modification factor $R_{\rm AA}$ and elliptic flow $v_2$, across both RHIC and LHC collision energies. This generalized framework lays a robust theoretical groundwork for high-precision heavy-flavor phenomenology, specifically facilitating future explorations into the baryon-rich regions mapped by the beam energy scan programs.

\begin{acknowledgements}
We thank Shanshan Cao, Hengtong Ding, and Rasmus N. Larsen for their discussions and valuable suggestions.
We also thank Yi Chen, Raymond Ehlers, Min He, Alexander Rothkopf, and Maria Lucia Sambataro for providing the relevant model data.
This work is supported by the National Natural Science Foundation of China (NSFC) under Grants No. 12375137, No. 12405138 and No. 12005114.
\end{acknowledgements}

\bibliography{Potential}

\end{document}